\documentclass{jors}
\usepackage[utf8]{inputenc}
\usepackage{graphicx}
\usepackage{colortbl}
\definecolor{mygray}{gray}{0.6}

\usepackage{color}
\usepackage{subfig}

\usepackage{commath}
\usepackage{bm}
\usepackage[sort, numbers]{natbib}

\usepackage{booktabs}
\usepackage{multirow}

\usepackage{listings}
\usepackage{xcolor}
\usepackage{tabularx}

\usepackage[normalem]{ulem}

\definecolor{codegreen}{rgb}{0,0.6,0}
\definecolor{codegray}{rgb}{0.5,0.5,0.5}
\definecolor{codepurple}{rgb}{0.58,0,0.82}
\definecolor{backcolour}{rgb}{0.95,0.95,0.92}

\lstdefinestyle{mystyle}{
    backgroundcolor=\color{backcolour},   
    commentstyle=\color{codegreen},
    keywordstyle=\color{magenta},
    numberstyle=\tiny\color{codegray},
    stringstyle=\color{codepurple},
    basicstyle=\ttfamily\footnotesize,
    breakatwhitespace=false,         
    breaklines=true,                 
    captionpos=b,                    
    keepspaces=true,                 
    numbers=none,                    
    numbersep=5pt,                  
    showspaces=false,                
    showstringspaces=false,
    showtabs=false,                  
    tabsize=2
}
\begin{document}






\vspace{0.5cm}

\section*{Title}
 {\tt simwave} - A Finite Difference Simulator for Acoustic Waves Propagation 

\section*{Paper Authors}

1. Souza, Jaime Freire de; \\
2. Moreira, João Baptista Dias;\\
3. Roberts, Keith Jared;\\
4. Gaioso, Roussian di Ramos Alves;\\
5. Gomi, Edson Satoshi; \\
6. Silva, Emílio Carlos Nelli; \\
7. Senger, Hermes.\\

\section*{Paper Author Roles and Affiliations}
{1. Federal University of São Carlos (UFSCar). Roles: Software, Validation, Writing - original draft, Writing - review \& editing.  \\
2. University of São Paulo (USP). Roles: Numerical methods, Software, Validation, Writing - original draft, Writing - review \& editing. \\
3. University of São Paulo (USP). Roles: Numerical methods, Software, Validation, Writing - original draft, Writing - review \& editing. \\
4. Federal University of São Carlos (UFSCar). Roles: Software, Writing - review \& editing.\\
5. University of São Paulo (USP). Roles: Supervision, Writing - original draft, Writing - review \& editing.\\
6. University of São Paulo (USP). Roles: Funding acquisition, Project administration, Supervision, Writing review. \\
7. Federal University of São Carlos (UFSCar). Roles: Funding acquisition, Project administration, Supervision, Writing - original draft, Writing - review \& editing. 
}

\section*{Abstract}


{\tt simwave} is an open-source Python package to perform wave simulations in 2D or 3D domains. It solves the constant and variable density acoustic wave equation with the finite difference method and has support for domain truncation techniques, several boundary conditions, and the modeling of sources and receivers given a user defined acquisition geometry. The architecture of {\tt simwave} is designed for applications with geophysical exploration in mind. Its Python front-end enables straightforward integration with many existing Python scientific libraries for the composition of more complex workflows and applications (e.g., migration and inversion problems). The back-end is implemented in C enabling performance portability across a range of computing hardware and compilers including both CPUs and GPUs.

\section*{Keywords}

Acoustic waves simulation, seismology, finite differences, high performance computing, Python.

\section{Introduction}


Acoustic waves are a means of energy propagation through a medium in space. These waves travel with a characteristic velocity and exhibit phenomena like diffraction, reflection and interference as they interact with the medium. The propagation of acoustic waves can be described by pressure variation, particle velocity, particle displacement, and/or acoustic intensity.
The propagation of acoustic waves is often used as a remote sensing tool to probe domains that are otherwise difficult to physically observe. Depending on the properties of the medium and the application, the simulation of acoustic waves may or may not consider variations in material density. For example, the acoustic wave equation with a constant density approximation is frequently used in seismic inversion workflows to estimate the P-wave velocity in the ground, which is later used to help locate raw material deposits such as oil and gas \cite{virieux2009overview,shearer2019introduction}. In medical imaging, similar methods are used that consider variations in material density or elasticity to study and diagnose tumors and other lesions in the human body \cite{Guasch2020, xia2017forward,mariappan2016magneto}. Acoustic tomography also plays an important role in understanding and monitoring ocean processes such as the global tides and internal waves \citep{munk1982observing, dushaw1997topex} and atmospheric turbulence \cite{keith1994acoustic}. In structural modeling, the acoustic wave can be used to identify failures in complex structures such as bridges and buildings \cite{shiotani2015visualization, li2017acoustic}. 

\indent Many wave propagators are part of comprehensive propriety codes that are developed by companies for industrial-grade workflows. In this context, usually the software is not available to independent researchers. Often many of these industrial workflows require computationally efficient implementations that can be used at many different computing scales, and this implies that re-implementation at some level is required.

\indent {\tt simwave} is a Python package that enables researchers to model acoustic waves propagation using short Python scripts with implementations that are verified and optimized for high performance. To be useful to a wide range of applications, the package is made to be flexible across hardware and software environments. Users interact with {\tt simwave} with a Python application programming interface (API) by passing user inputs that control the desired accuracy of the simulation. Many components of {\tt simwave} are implemented for applications with geophysical exploration and the simulation of waves can occur with either the assumption of constant or a variable density medium.

\subsection{Applications}
The acoustic wave is often used to solve inversion problems to estimate material properties such as in full waveform inversion (FWI) \citep{fichtner2010full}. These inverse problems are particularly computationally demanding as they require many wave propagation simulations in order to produce meaningful solutions to the inverse problem. As a result, the primary computational cost of the inversion process is proportional to the speed at which one can simulate the propagation of a wave.

\indent An overview of a typical inversion setup is shown in Figure~\ref{fig:overview}:
\begin{figure}
    \centering
    \includegraphics[scale=0.75]{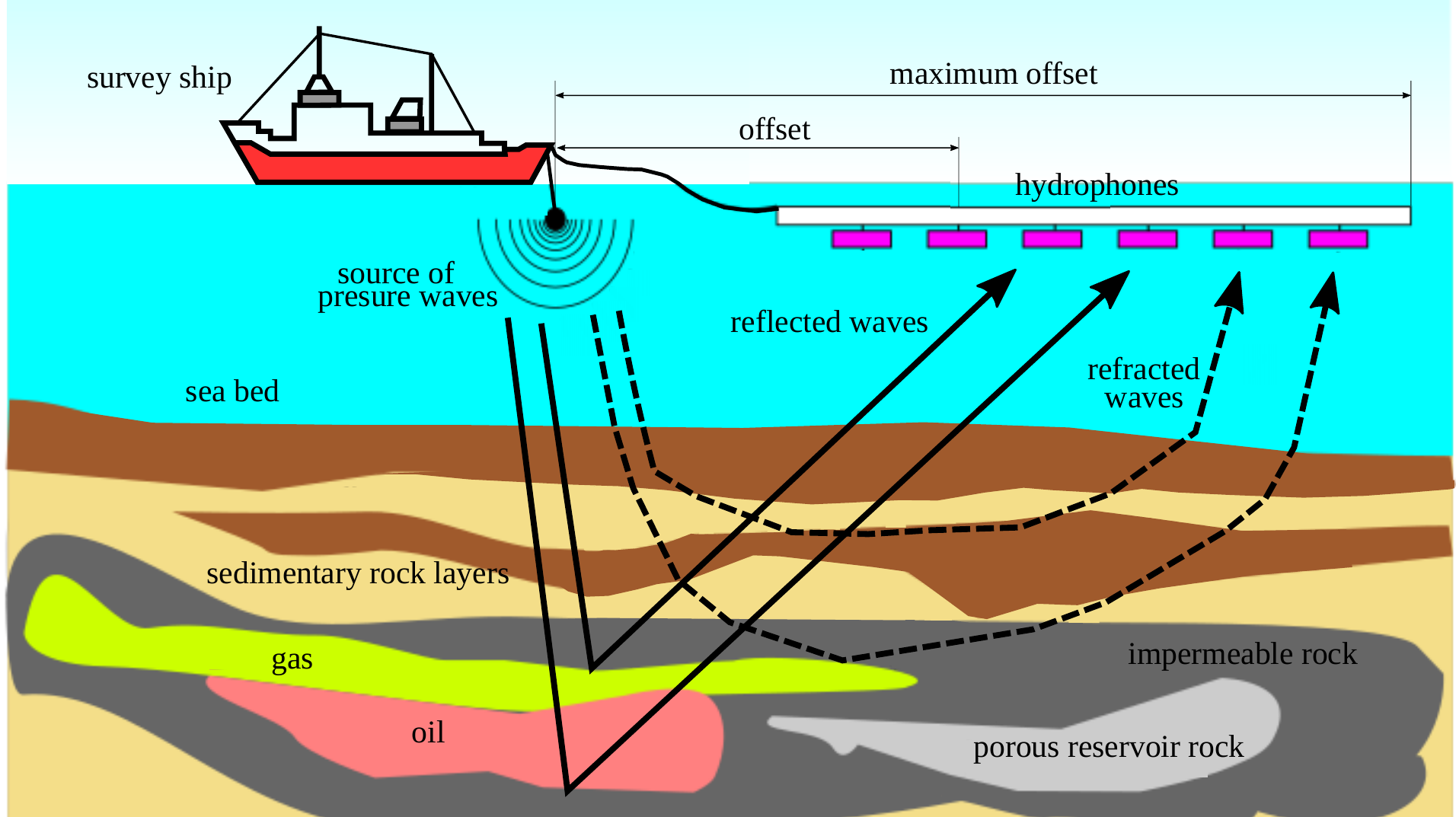}
    \caption{An application of acoustic wave propagation as a remote sensing tool.}
    \label{fig:overview}
\end{figure}
As an example, in a typical FWI setup in a marine environment, a ship tows a cable with hundreds of recording devices termed \textit{receivers} potentially several kilometers long. On the ship, small controlled explosions known as a \textit{shots} or \textit{sources} are periodically fired. These sources propagate acoustic waves that interact with the subsurface medium and produce signals recorded by the receivers. The collection of seismic signals for a particular source explosion event is referred to as a \textit{shot record} and the quantity and the location of the sources with respect to the location of the receivers is refereed to as an \textit{acquisition geometry}. 
A similar technique is applied to model how ultrasound energy is transmitted through the skull to generate accurate three-dimensional images of the human brain with sub-millimeter resolution \cite{guasch2020full}.





\section{Governing equations}

 The propagation of mechanical waves can be modeled with the elastic wave equation \cite{fichtner2010full}:
\begin{equation}
    \rho(\mathbf{x}) \frac{\partial^{\; 2} \mathbf{u}}{\partial t^2}(\mathbf{x}, t) = \nabla \cdot \bm{\sigma}(\mathbf{x}, t) 
    + \rho(\mathbf{x}) \mathbf{b}(\mathbf{x}, t)
    \label{eq:motion}
\end{equation}
where $\mathbf{u}$ is the particle displacement vector, $\bm{\sigma}$ is the stress tensor, $\rho$ is density, $\mathbf{b}$ corresponds to external body forces, and 
$\nabla = (\frac{\partial}{\partial x}, \frac{\partial}{\partial y}, \frac{\partial}{\partial z})$ in cartesian coordinates. 
Vectors and tensors are denoted with bold letters. scalars are Equation~\eqref{eq:motion} is derived through
the conservation of linear momentum. The propagation of elastic waves leads to longitudinal (P) waves, transversal (S) waves, P-to-S wave conversions, besides free surface phenomena such as Rayleigh and Love waves \cite{atkin2005introduction}. 

\indent For a linear elastic non-dissipative medium, the relationship between stresses and 
strains $\mathbf{\varepsilon}$ is given by 
$\sigma = \mathbf{C}:\bm{\varepsilon}$. The fourth order 
elastic tensor $\mathbf{C}$ has between $2$ up to $21$ variables depending on the degree of anisotropy of the materials being considered \cite{virieux2009overview}.
The complexity of $\mathbf{C}$ influences computational cost. For instance, an efficient finite difference implementation of a wave propagator for a relatively large problem ($768^3$ DoFs) considering transverse  isotropic medium is about five times slower than compared to an isotropic medium  \cite{luporini2020architecture}.

\indent An often adopted alternative \cite{igel2017computational} is to model the P-wave propagation using the acoustic wave equation, which is obtained by assuming an isotropic medium and neglecting shear strains:

\begin{equation}
    \frac{1}{\kappa(\mathbf{x})} \frac{\partial^{\; 2} p}{\partial t^2}(\mathbf{x}, t) - \nabla \cdot 
    \left( \frac{1}{\rho(\mathbf{x})}\nabla p (\mathbf{x}, t) \right) = - \nabla \cdot \mathbf{b}(\mathbf{x}, t) 
        \label{eq:acoustic_density}
\end{equation}
where $\kappa$ is the bulk modulus relating scalar pressure $p$ and displacement $\mathbf{u}$ via the expression $p = - \kappa \nabla \cdot \mathbf{u}$. If the density varies significantly slower than the pressure field,
Equation~\eqref{eq:acoustic_density} can be simplified by making the assumption of constant density in the medium to:
\begin{equation}
    \frac{\partial^{\; 2} p}{\partial t^2}(\mathbf{x}, t) - c^2(\mathbf{x})\nabla ^ 2 p (\mathbf{x}, t) 
    = - \rho  \; c^2(\mathbf{x}) \nabla \cdot \mathbf{b} (\mathbf{x}, t) 
        \label{eq:acoustic}
\end{equation}
where $c = \sqrt{\kappa / \rho}$ is the wave speed.

\indent Equations~\ref{eq:acoustic_density} and ~\ref{eq:acoustic} are frequently used in active source seismic imaging \citep{fichtner2010full}. Despite not representing the full complexity of the propagation of waves, the acoustic wave equation can still suffice. For one, not all data acquisition equipment can effectively capture or utilize more complex wave propagation physics. Secondly, solving a scalar partial differential equation (PDE) \eqref{eq:acoustic} is considerably computationally cheaper and requires less run-time memory than the vectorial PDE required by the elastic wave equation \eqref{eq:motion}. 

\indent If the wave propagation constitutes a step of an imaging workflow, the number of distinct material parameters is also relevant to computational cost. For example, while the acoustic approximation for constant density \eqref{eq:acoustic} can be defined in terms of the wave speed $c$, the wave equation with varying density \eqref{eq:acoustic_density} needs the inversion of two independent fields: $\rho$ density and P-wave velocity.

\indent In this work the acoustic wave equation in its $2^nd$ order form with either constant \eqref{eq:acoustic} or variable density \eqref{eq:acoustic_density} was discretized using the finite difference method. Both the constant and variable density finite difference stencils' accuracy goes to up to $20^{th}$ order in space and can be controlled at run time by the user. A second order central finite difference approximation is employed for the time derivative to create an explicit time-stepping scheme.

\subsection{Boundary conditions and domain truncation} 

The application of boundary conditions and domain truncation techniques play an important role in the simulation of the acoustic wave. For example, applications such as non destructive testing and medical imaging workflows often need to enforce Dirichlet boundary conditions to emulate a free-surface. Seismic applications often need to damp simulated waves from reflecting off domain boundaries. In these cases, domain truncation methods can be used to effectively absorb outgoing waves from the interior of a computational region without reflecting them back into the interior but at the cost of additional terms in the governing equations.

\indent In most acoustic wave applications, a combination of an absorbing boundary condition \cite{engquist1977absorbing} and a domain truncation technique like an absorbing boundary layer are used. Occasionally, a special treatment is also required to represent the free-surface boundary to model reflections \cite{doi:10.1190/1.1444107}. These boundary condition techniques range from enforcing Robin boundary conditions \cite{clayton1977absorbing, higdon1987numerical} to more complex approaches that involve modifying the acoustic wave equation and augmenting the physical domain \cite{berenger1994perfectly}.

\indent {\tt simwave} currently supports both Neumann and Dirichlet boundary conditions, which can be used on any number of the domain boundaries in addition to a user-configurable absorbing boundary layer (ABL) \cite{gao2017comparison}. In the case of an ABL, the domain becomes $\Omega = \Omega_{0} \bigcup \Omega_{ABL}$ where $\Omega_{0}$ is the physical domain and $\Omega_{ABL}$ is the additional layer of user-defined width to absorb outgoing waves. In the case of the ABL, a non-zero damping term $\eta$ is added to the original wave equation within $\Omega_{ABL}$:
\begin{equation}
     \frac{\partial^{\; 2} p}{\partial t^2}(\mathbf{x}, t) + 2 \eta \frac{\partial p}{\partial t}(\mathbf{x}, t) - c^2(\mathbf{x})\nabla ^ 2 p (\mathbf{x}, t) 
    = - \rho  \; c^2(\mathbf{x}) \nabla \cdot \mathbf{b} (\mathbf{x}, t) 
    \label{eq:acoustic_damped}
\end{equation}
where $\eta$ is zero everywhere except in the ABL. In the ABL, $\eta = \alpha d(\mathbf{x})^p$ in which $\alpha$ and $p$ are two parameters that control the profile of the damping function, while $d(\mathbf{x})$ is the shortest distance from $\mathbf{x}$ to the $\Omega_{0}$.

\subsection{Time discretization}

By multiplying Eq. \eqref{eq:acoustic_density} by the density $\rho$ and expanding the expression under the divergent operator, the acoustic
wave equation for variable density (with damping) at the instant $t=t_n$ may be written as:
\begin{equation}\label{eq:open_acoustic_density}
    \frac{1}{c^2} \left( \frac{\partial^{\; 2} p}{\partial t^2}(t_n) + 2 \eta \frac{\partial p}{\partial t}(t_n) \right) + \frac{\nabla\rho}{\rho} \cdot \nabla p(t_n)
    - \nabla^2 p(t_n) = f(t_n) 
\end{equation}

\indent The time axis is discretized uniformly such that  $t_n = n \Delta t$ for $n=0, \ldots, N$ under a certain time step size $\Delta t$. 
A second-order accurate in time central finite difference scheme is chosen to approximate the time derivatives
\begin{equation}\label{eq:central}
    \frac{d p}{dt}(t_n) \approx \frac{p^{n+1} - p^{n-1}}{2 \Delta t}, \quad \frac{d^2 p}{dt^2}(t_n) \approx \frac{p^{n+1} - 2 p^{n} + p^{n-1}}{\Delta t^2}.
\end{equation}
In that case, an explicit time stepping scheme is obtained:
\begin{equation}\label{eq:explicit}
    p^{n+1} = \frac{c^2 \Delta t^2}{1 + \eta \Delta t}( \nabla^2 p^n - \frac{\nabla \rho}{\rho} \cdot \nabla p^n) + 2p^n - (1 - \eta \Delta t) p^{n-1} 
\end{equation}

\indent In practical applications,
it remains important to be able to automatically determine a numerically stable timestep for the discretization. For the second-order timestepping method used in this work, the necessary condition to select a numerically stable timestep $\Delta t$ is given by \cite{Lines}: 
\begin{equation}
    \Delta t \leq \frac{2 \, \Delta x}{c_{max} \sqrt{a}}
    \label{eq:cfl}
\end{equation}
in which $c_{max}$ is the maximum seismic velocity in the domain, $\Delta t$ is the maximum timestep that can remain numerically stable,  $\Delta x$ is the grid spacing, and $a$ is the sum of the finite difference coefficients involved with the spatial derivative terms in the wave equation. Note that $a$ considers the usage of effect of arbitrarily higher order stencils for space derivative terms.

\indent The timestepping scheme was implemented in a way such that wave propagators only need to keep in memory at most two time levels simultaneously, which reduces run-time memory load.

\subsection{Space discretization}

The computational domain is discretized with a regular grid with uniform spacing $\Delta x_i$ in each axis $x_i$, where $i$ goes from $1$ up to $2$ in 2D and $3$ in 3D.

\indent The spatial derivatives are approximated by central finite differences of even spatial orders up to 20. Along the $x_i$ axis, the first and second derivatives at $x_i=k$ read as:
\begin{align}
        \partial_{x_ix_i} \phi_k = \frac{1}{\Delta x_i^2} ( v_{0} \phi_k + \sum^{r}_{j=1} v_{i} ( \phi_{k+j} + \phi_{k-j} )) \label{eq:spatial_discretization_1} \\
        \partial_{x_i} \phi_k  =  \frac{1}{2\Delta x_i} ( \sum^{r}_{j=1} w_{i} ( \phi_{k+j} - \phi_{k-j} ))  
        \label{eq:spatial_discretization_2}
\end{align}
in which $v_i$ are the coefficients of even spatial order for central finite difference schemes for second-order derivatives, $w_i$ are the weights for central finite different schemes for even spatial order for first-order derivatives, and $r$ represents the stencil radius. The fully discretized stencil is obtained by substituting the expressions from \eqref{eq:spatial_discretization_1} and \eqref{eq:spatial_discretization_2} into Eq. \eqref{eq:explicit}. At the boundary, the domain is augmented with a number of ghost nodes that depends on the order of the stencils used to discretize the spatial derivatives.

\subsection{Sources and receivers}
\label{sec:sources_receivers}

The approach detailed in \citep{Hicks2002} is used to implement a body force at an arbitrary location within the grid and also to interpolate wave field solutions to receiver locations. Briefly, the source term is given by:
\begin{equation}
    f_n = S \, d_n = S \, [W(n + \alpha) \,sinc(n + \alpha)]
    \label{eq:source_impl}
\end{equation}
in which $-0.5 < \alpha \leq 0.5$ and $n$ represents an integer denoting a grid point, $d_n$ represents a band-limited spatial delta function, and $S$ is a time-varying wavelet. 

\indent The band-limited spatial delta function $d_n$ is represented using a Kaiser window. The window function $W$ given by:
\[
    W(x)= 
\begin{cases}
   \frac{ I_0 (b \sqrt{1 - (x/r)^2})}{I_{0}(b)},& \text{for } -r \leq x \leq r\\
    0,              & \text{otherwise}
\end{cases}
\]
with the one free parameter $b$ associated with the window, the half-width of the filter $r$, and $I_0$ is the zeroth-order Bessel function of the first kind. Optimal values for $b$ from \citep{Hicks2002} are programmed for wavenumbers $k_{max} = \frac{1}{2}\pi$ given varying $r$. Ideally, the value of $r$ should be kept as low as possible; however, this depends on the application and the desired numerical accuracy. With that said, the user can specify the desired value for $r$.

\begin{figure}[ht]
    \centering
    \includegraphics[width=\textwidth]{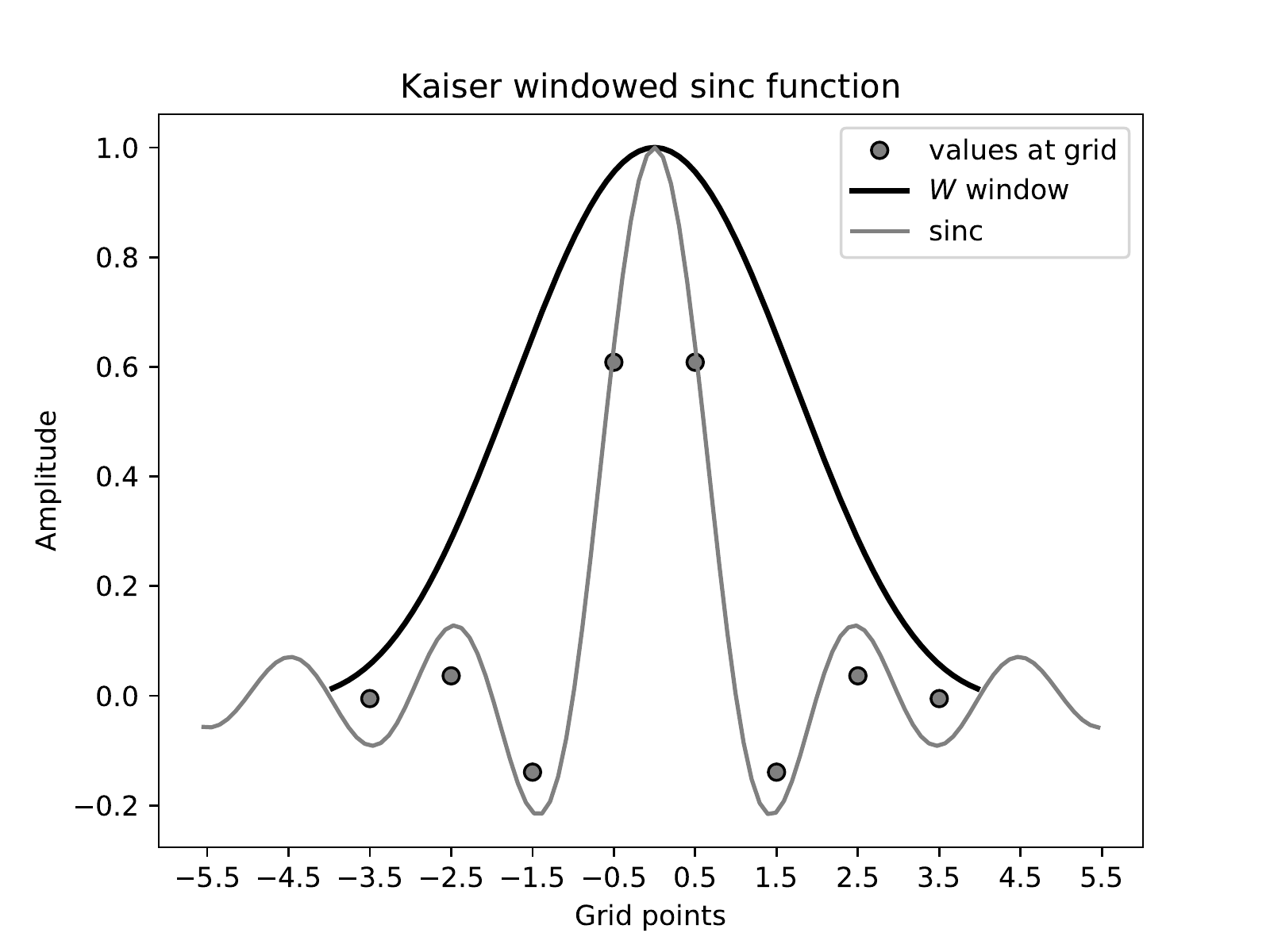}
    \caption{Windowing for a point at a distance of $0.5$ grid points from its neighbors.}
    \label{fig:kaiser}
\end{figure}

\indent Figure \ref{fig:kaiser} displays an example of a Kaiser Window $W$ together with a $sinc$ 
function and the corresponding weights multiplying the grid point values. The source (or receiver) is at a distance of $0.5$ points from its neighbors, $b=6.31$ and $r=4$ in this instance.

\indent It similarly follows that the wave field solution $p_n$ can be recorded to a set of arbitrary receiver locations $R$ in either 2D or 3D through: 
\begin{equation}
    R = \sum_{n=-r}^{r} p_n \, d_n
\end{equation}

\indent {\tt simwave} permits the user to define an arbitrary time-varying wavelet $S$. By default, a function to generate a time-varying Ricker wavelet for a user-specified peak frequency is implemented.

\subsection{Verification of numerical implementation}
\label{sec:verification}


In order to verify that the numerical solutions produced by {\tt simwave} are mathematically correct, we conduct several convergence tests in which we compare the order of accuracy of the discretized wave equation against theoretical values.

\begin{figure}[ht]
    \centering
    \includegraphics[width=\textwidth]{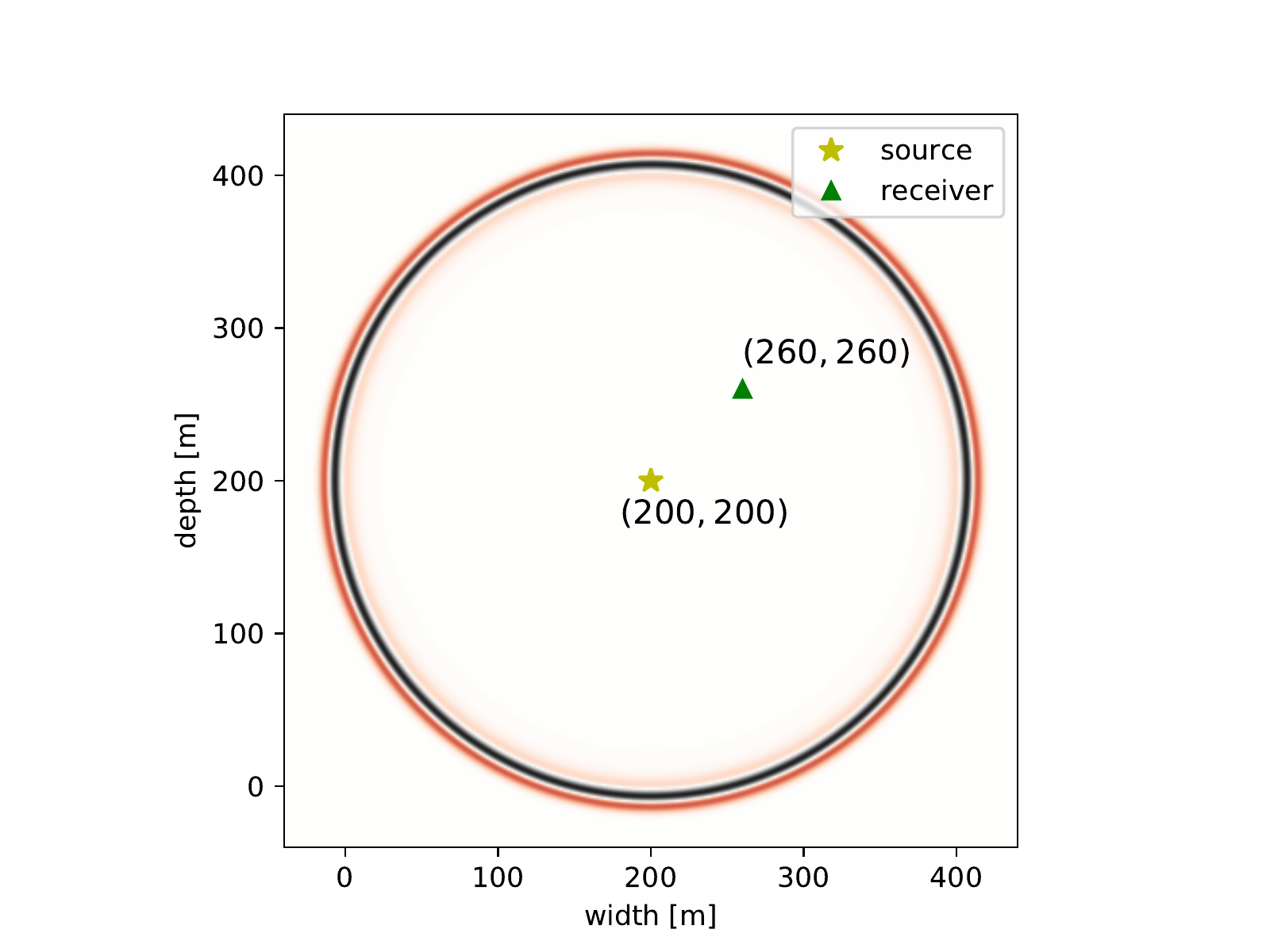}
    \caption{Simulation setup for the verification of the acoustic wave equation implementation.}
    \label{fig:conv_setup}
\end{figure}

\indent A domain of $400$ $\times$ $400$ meters consisting of a homogeneous velocity model with $c = 1.5$ km/s is considered. At the center of the domain, a point source with a time varying signal $s(t)$ produces a wavefield $u(\mathbf{r}, t)$, where $\mathbf{r}$ denotes the distance from the source. A receiver at a distance of approximately 85 meters from the source registers the wave amplitude for $t = 150$ microseconds (Figure \ref{fig:conv_setup}). The wave at the final instant $t = 150$ ms is also plotted, showing that the wave front never reaches the computational boundary. A Kaiser window width of 4 points is used both for source injection and receiver value interpolation. Wave and velocity field, as well the values collected at receivers are represented as single precision floating point numbers.

\indent Numerical solutions are compared to an analytical solution \cite{watanabe2015green} given by: 
\begin{equation}
    u(\mathbf{r}, t) = - \frac{i}{2} \int_{-\infty}^{\infty} H_0^{(2)} 
    \left(\frac{\omega}{c} \mathbf{r} \right) 
    \hat{s}(\omega) e^{i \omega t} \dif \omega 
    \label{eq:analytical_solution}
\end{equation}
where $H_0^{(2)}$ is the Henkel function of second kind and $\hat{s}$ is the Fourier transform of the original signal $s$. This analytical solution is valid as long as the source is punctual and boundary effects can be ignored.

\begin{figure}[ht]
    \centering
    \includegraphics[width=\textwidth]{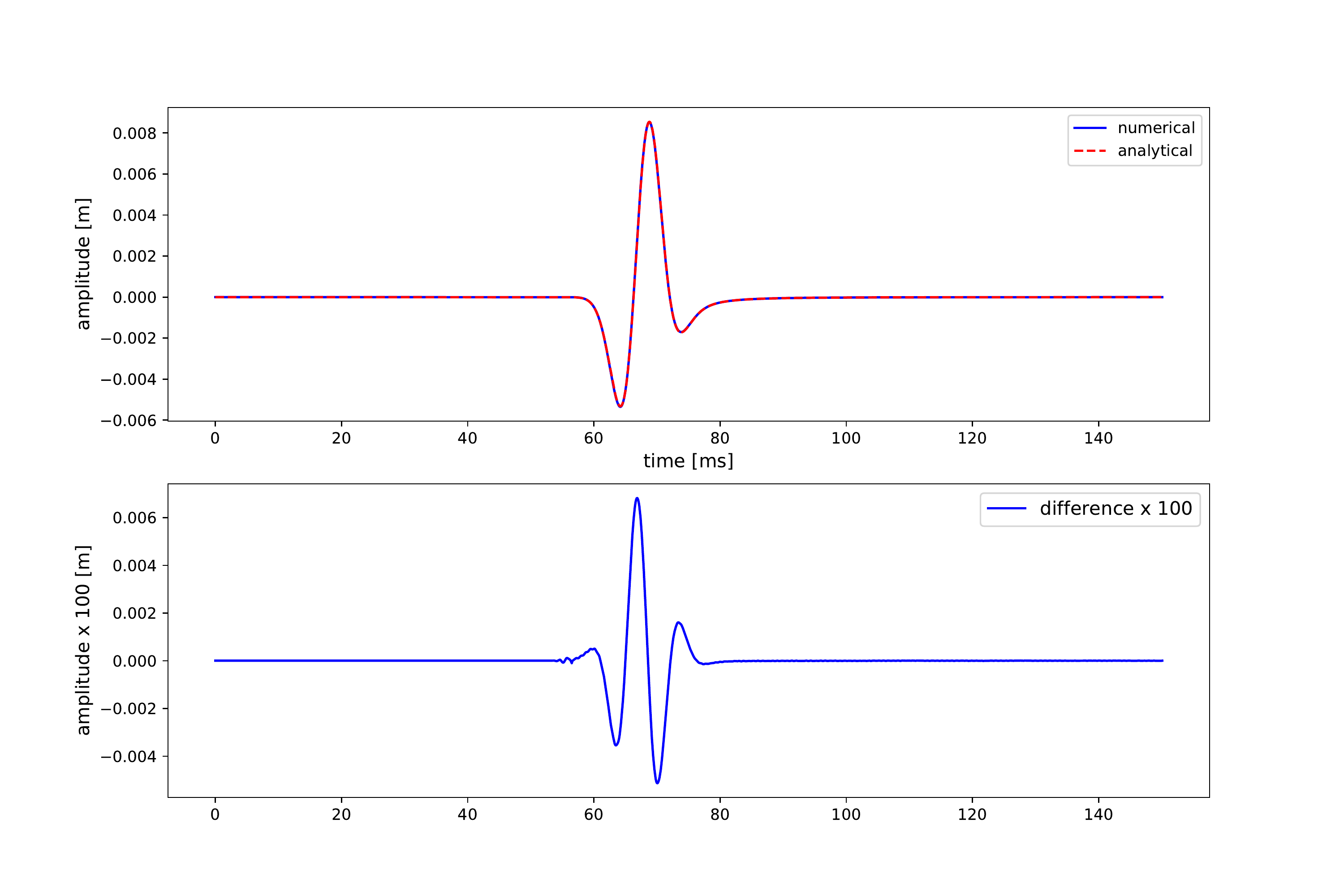}
    \caption{Comparison between numerical and analytical solution.}
    \label{fig:analytical}
\end{figure}

\indent The domain is discretized as a square grid with spacing of 0.5 meters between nodes along both axes, and the time axis is discretized with a timestep of
0.1 ms. As shown in Fig. \ref{fig:analytical}, the numerical solution is able to reasonably approximate the analytical one, as their difference is two orders of magnitude lower than the amplitudes at the receiver.

\begin{figure}[ht]
    \centering
    \includegraphics[scale=0.8]{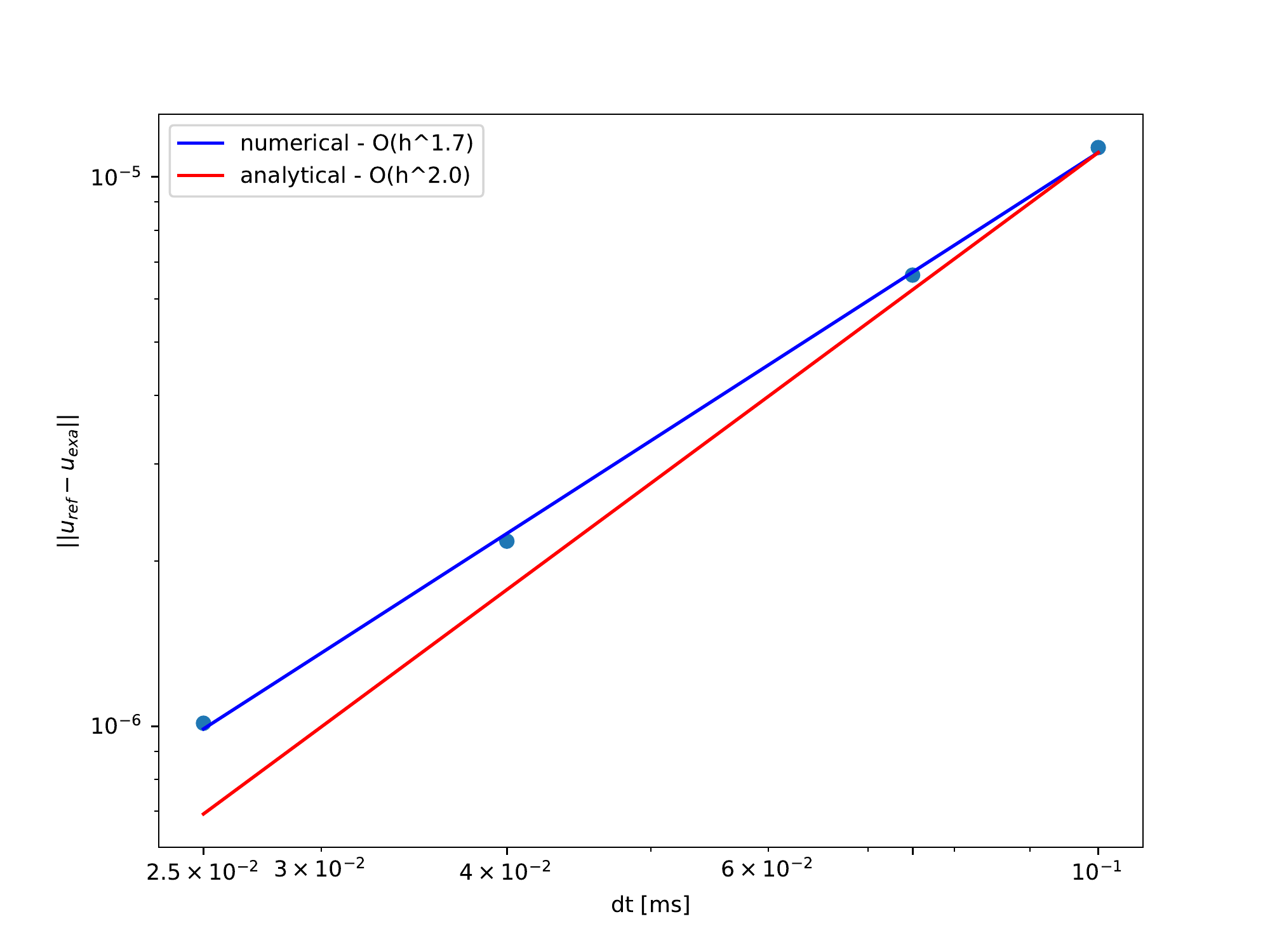}
    \caption{Numerical accuracy rate in time.}
    \label{fig:time_comparison}
\end{figure}

\indent In order to verify the time discretization, we fix the spatial grid with spacing $h=0.5$ meters and evaluate the 
Euclidean norm of the difference between the numerical solution $u_{ref}$ and the exact solution $u_{exa}$ at the receiver location. Since the time finite difference stencil employed is of second order, the error should decrease to the second order as $O(\Delta t^2)$. Fig. \ref{fig:time_comparison} displays the convergence rate alongside the theoretical curve, which demonstrates good agreement between theoretical and observed values.

\begin{figure}
    \centering
    \includegraphics[scale=0.8]{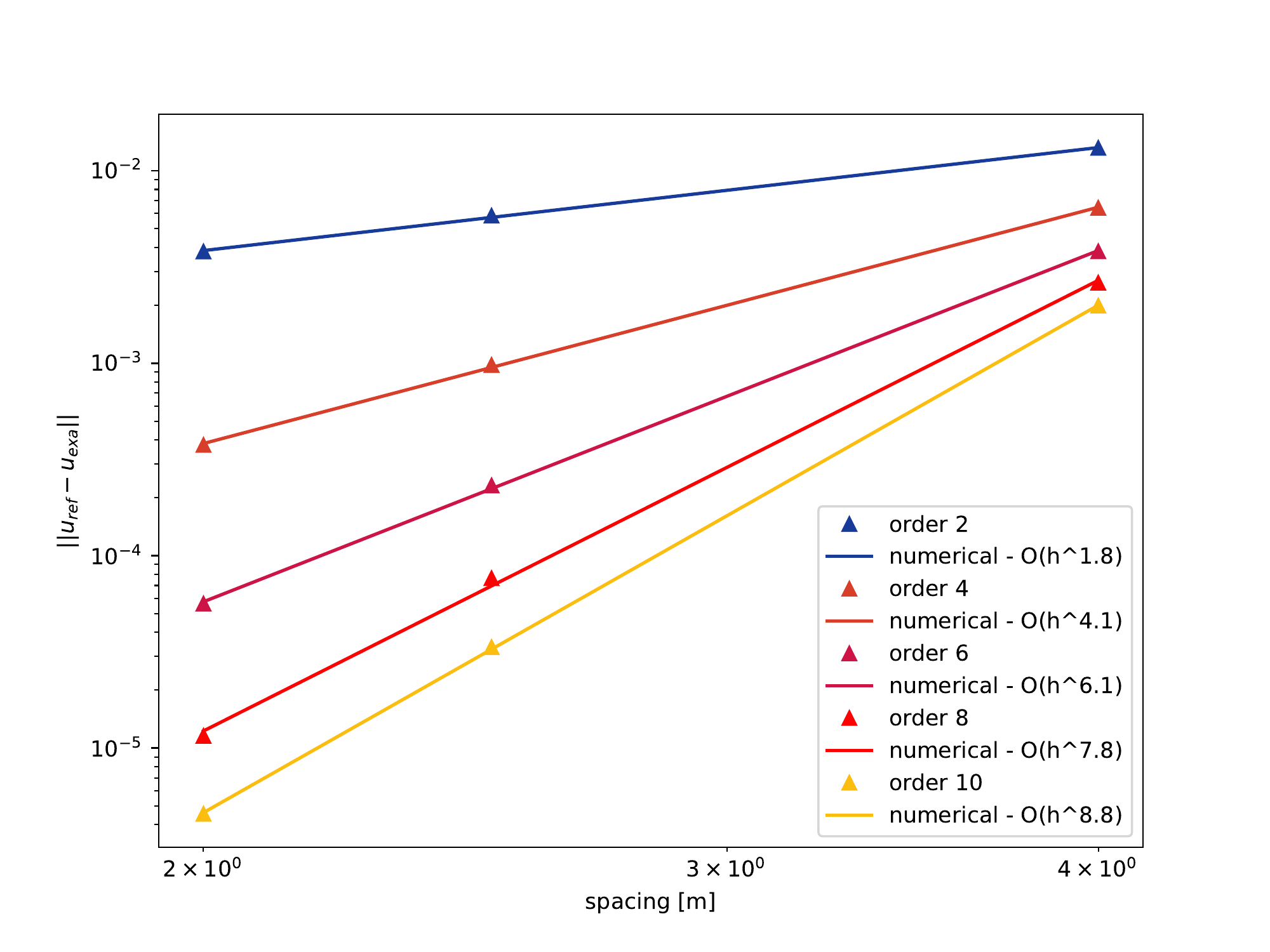}
     \caption{Numerical accuracy rate in space.}
    \label{fig:space_comparison}
\end{figure}

\indent A similar analysis is performed regarding the space discretization. For a sufficiently small and fixed $dt = 0.025$ ms to minimize the influence of time discretization error, the spatial error is evaluated for different values of grid spacing $h$. Stencils with spacial orders up to $10$ are considered and results are shown in Fig. \ref{fig:space_comparison}. The convergence rates agree well with theoretical values for orders up to $8$, and start to diverge from it as the magnitude of the spatial error becomes of the same magnitude as the time discretization error.

\begin{figure}
    \centering
    \includegraphics[scale=0.85]{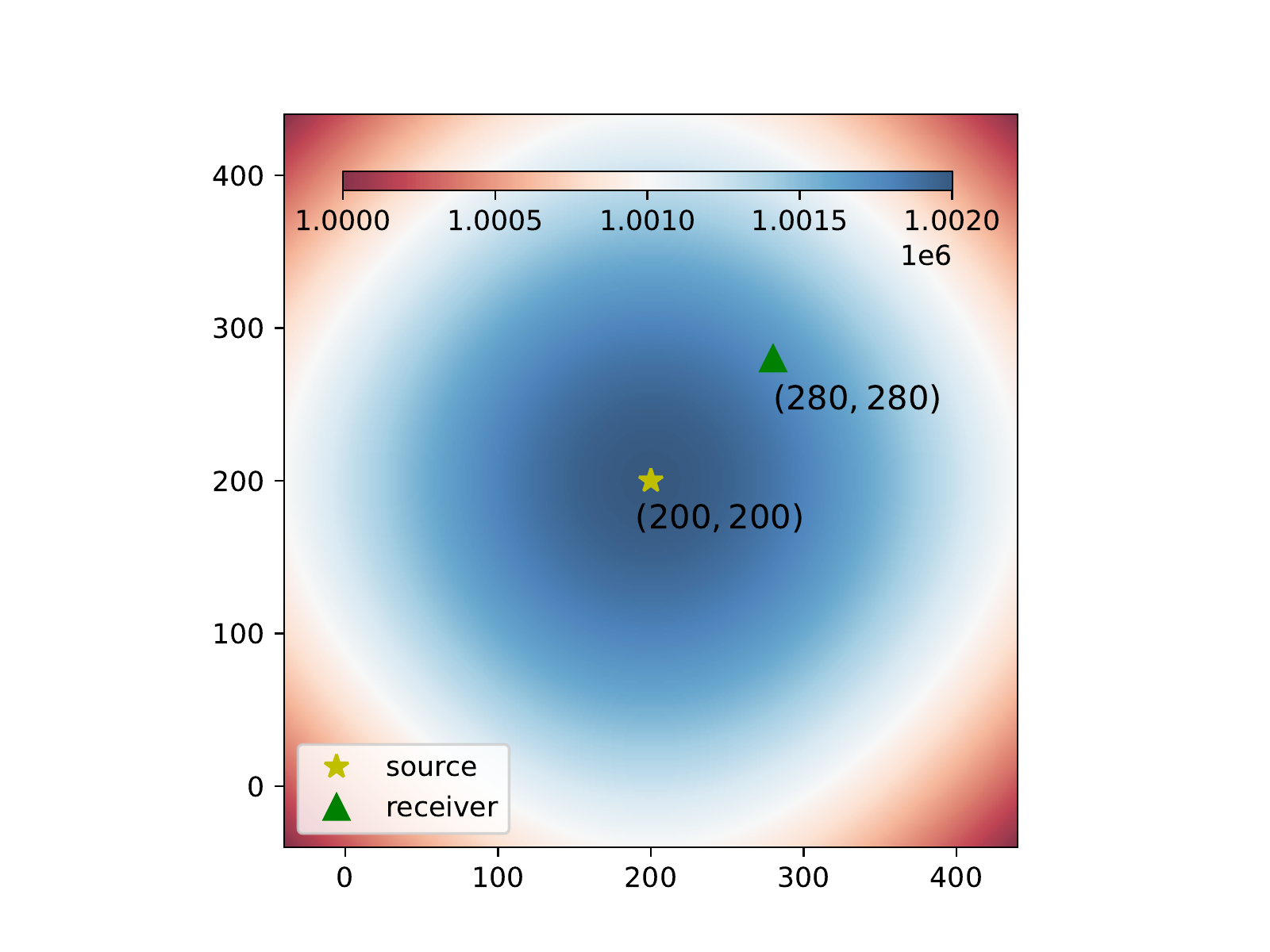}
     \caption{Density distribution used in the verification of the acoustic wave equation with variable density. The position of the source/receiver pair is also shown.}
    \label{fig:density_field}
\end{figure}

\begin{figure}[t]
    \centering
    \subfloat[Comparison for the whole time interval.]{
        \includegraphics[scale=0.45]{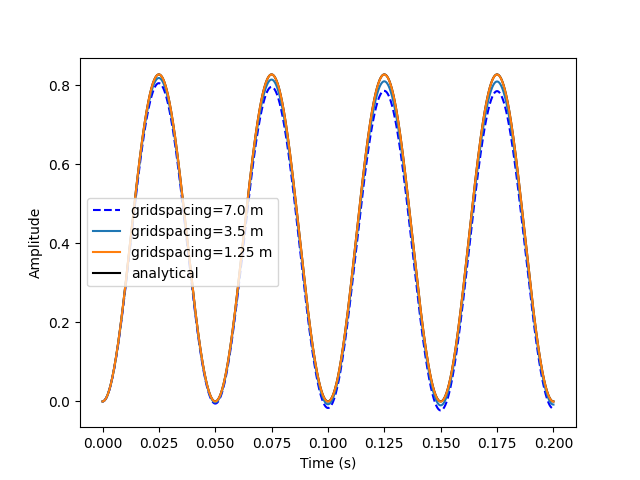}
        \label{fig:mms}
    }
    \subfloat[At a single peak.]{
        \includegraphics[scale=0.45]{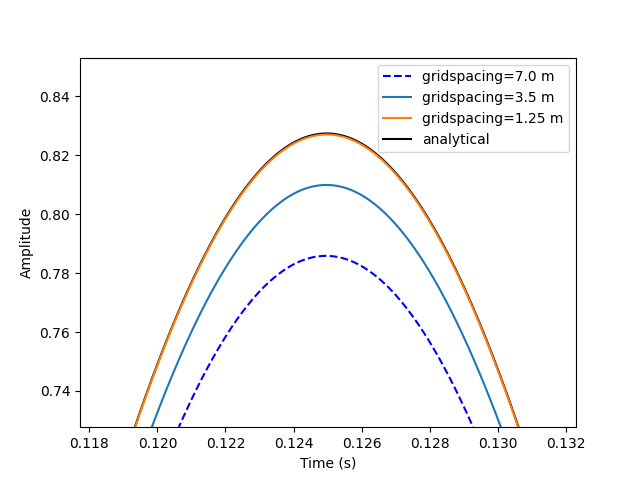}
        \label{fig:mmsb}
    }
    \caption{Comparison between numerical and analytical solution for variable density at the receiver location (280, 280).}
    \label{fig:mms_comparison}
\end{figure}

\indent Finally, the wave equation with variable density is verified by the Method of Manufactured Solutions (MMS) \cite{salari2000code}.
A domain of $440$ $\times$ $440$ meters consisting of a homogeneous velocity model with $c = 2 $km/s is considered. 
A point source at the center of the domain produces a wavefield $u^*(x, z, t)$, where $x, z$ are Cartesian coordinates. A receiver at a distance of approximately 113 meters from the source registers the wave amplitude for $t = 200$ microseconds
as displayed in Figure \ref{fig:density_field}.
The time interval is discretized with a timestep of 0.05 ms.
All relevant fields are represented with double precision floating point numbers.
In order for the use of the variable density equation to be meaningful, a spatially varying density field $\rho$ is chosen:
\begin{equation}
    \rho = \left( 1000 + \sin(\frac{\pi}{440} x) \right) 
           \left( 1000 + \sin(\frac{\pi}{440} z) \right)
\end{equation}
\begin{figure}
    \centering
    \includegraphics[scale=0.7]{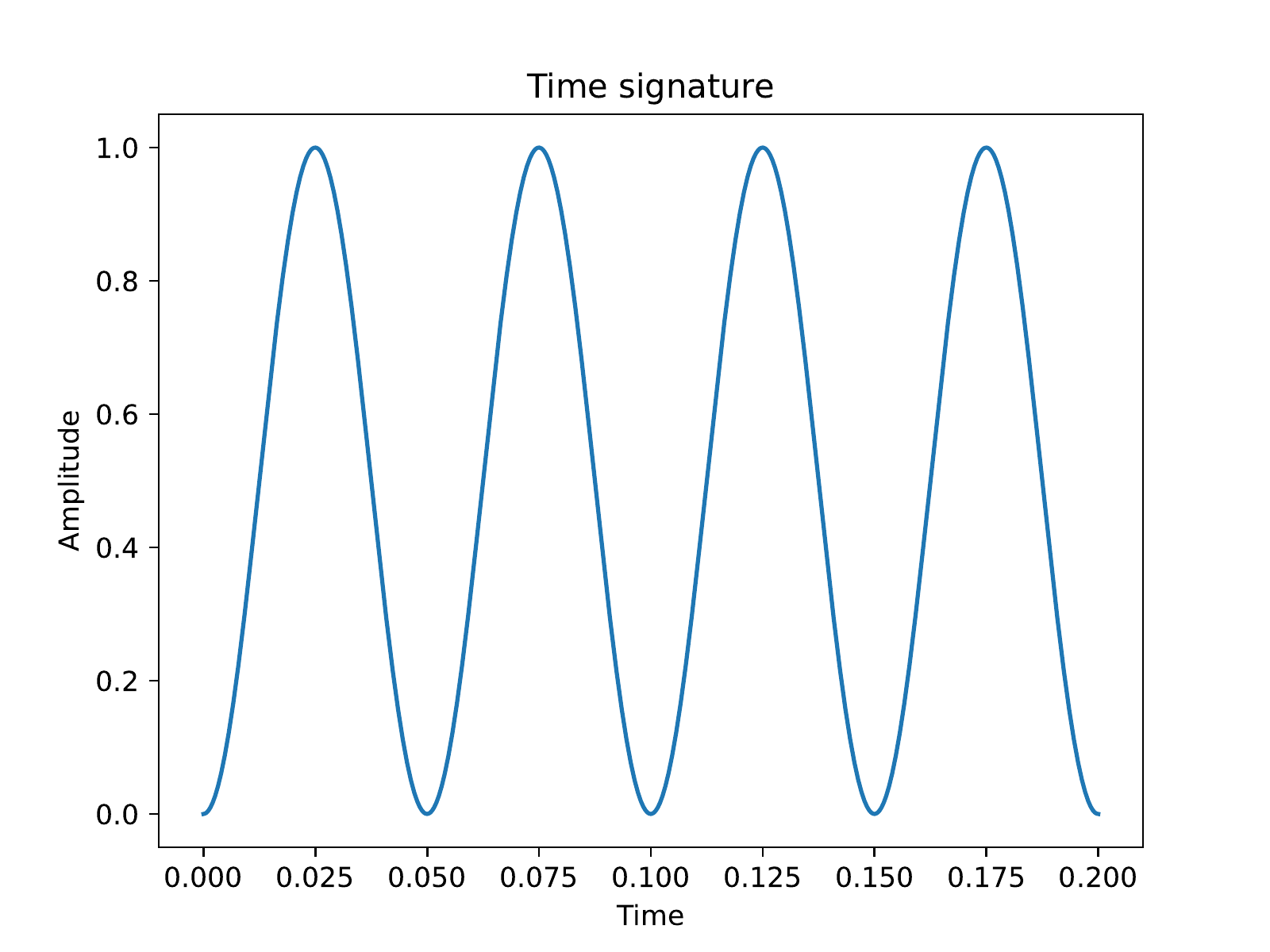}
    \caption{Time dependency of the solution $u^*$ used in the verification of the acoustic wave equation with variable density.}
    \label{fig:time_signature}
\end{figure}
The MMS consists in deriving the forcing term and boundary conditions for a PDE from a given solution. The following field is chosen as the ansatz:
\begin{equation}
    u^*(x, z, t) = \sin(\frac{\pi}{440} x) \sin(\frac{\pi}{440} z) \sin(20 \pi t) \sin(20 \pi (t + dt)).
    \label{eq:mms}
\end{equation}
The density field $\rho$ is plotted in Figure \ref{fig:density_field}. The solution $u*$ has the same spatial dependency as the density, 
while the time dependency is shown in Figure \ref{fig:time_signature}.
The appropriate forcing is derived by direct substitution into Eq. \eqref{eq:acoustic_density}. 
One can also verify that $u^*$ satisfies Dirichlet boundary conditions. The dependency in time is so that
$u^*$ is zero for the two first time steps $t=0$ and $t=dt$. Figure \ref{fig:mms_comparison} displays a comparison between analytical
and numerical solutions for several grid spacing values. The numerical solution seems reasonably able to approximate the analytical solution,
since as the grid spacing gets smaller, the numerical solution approaches the expected theoretical values.


\section{Code architecture and implementation}

\begin{figure}
    \centering
    \includegraphics[scale=0.3]{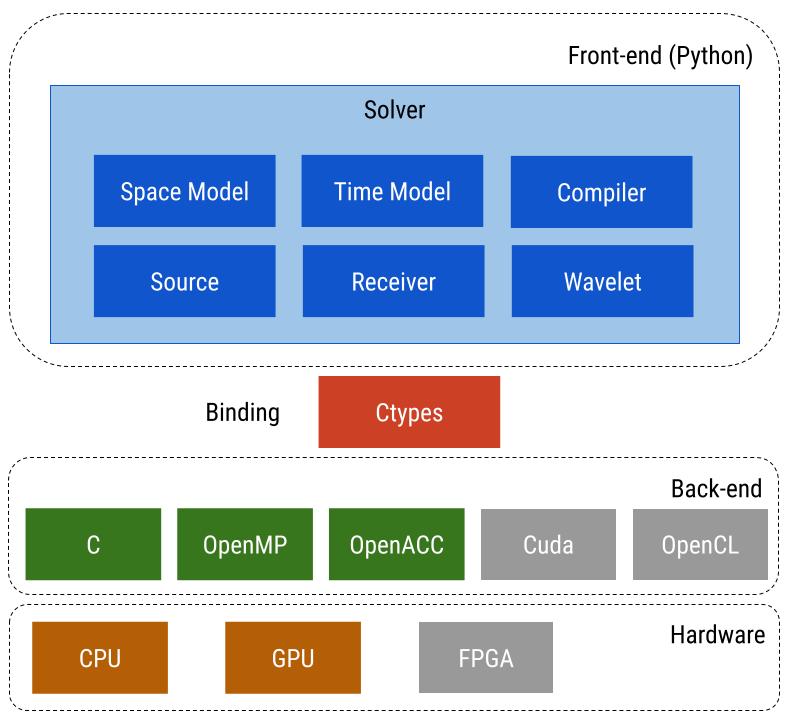}
    \caption{The architecture of {\tt simwave}: the components in gray will be implemented in future work.}
    \label{fig:simwave-architecture}
\end{figure}

For better separation of concerns, the architecture of {\tt simwave} is organized into two layers (Figure~\ref{fig:simwave-architecture}). 
A Python front-end is implemented to provide a user-friendly interface which facilitates application development and integration with other scientific software libraries such as SciPy \cite{virtanen2020scipy} and many others. 
A minimum body of knowledge is required from the application developer, for choosing a back-end, a compiler and its flags. 
All parallel processing strategies and hardware specific optimizations are implemented in the back-end.
The performance critical components are implemented in the back-end which is written in ANSI C (sequential), or in C plus some support for parallelism (e.g., OpenMP, OpenACC, etc). The integration between the front-end and back-end uses Ctypes. 

\subsection{The front-end}


The front-end provides the Python classes and functions with intuitive design for domain application programmers. 
The simulation of a wave propagation is performed by configuring and instantiating a {\tt Solver} object. 
The solver aggregates a set of objects that encapsulate important simulation parameters including:

\begin{itemize}
    \item {\tt SpaceModel:} This class defines the domain as a 2D or 3D axis-aligned regular Cartesian grid and requires additional numerical parameters to specify the spatial discretization. It configures the spatial order of the finite difference stencil. Boundary conditions and absorbing layers are also enabled by calling its method {\tt config\_boundary()}.  The {\tt SpaceModel} class requires grid-shaped dataset containing scalar values for all the grid points. For example, in seismology the seismic velocity values are typically supplied, while the spatially variable density is optional.
  
    \item {\tt TimeModel:} Objects of this class encapsulate temporal discretization parameters for the wave simulation, such as the start time, end time, and the timestep. This class can automatically calculate a numerically stable simulation timestep $\Delta t$ that respects the CFL conditions \cite{Lines} from a {\tt SpaceModel} object. The user can optionally specify $\Delta t$ if needed.

    \item {\tt Source:} This class implements source injection as described in Section \ref{sec:sources_receivers} according to the quantity and their locations in the domain provided by the programmer. Notice that multiple sources can be enforced simultaneously.
   
    \item {\tt Receiver:} similar to the {\tt Source} class, the {\tt Receiver} represents a set of receivers positions across the domain. These receivers represent recording devices (e.g. hydrophones) that record wave signals and can be used to generate seismograms for the simulation.
    
    \item {\tt Wavelet:} This class represents a time varying wavelet to be injected into the domain. The user can specify a custom call-back function that describes the variation in time of the body force. {\tt simwave} also provides a {\tt RickerWavelet} default sub-class which extends the {\tt Wavelet} and implements a Ricker wavelet.
    
    \item {\tt Compiler:} This object encapsulates  compilation parameters for the generation of C code, such as the compiler implementation (e.g. gcc, icc, clang) and compiler flags. The {\tt Compiler} class is responsible for compiling the C code and generating a shared object at run time. Despite belonging to the front-end stack, this object is used to generate the back-end code.

\end{itemize}

\subsection{The back-end}

The back-end layer, which  solves the PDEs and simulate the propagation of acoustic waves, is implemented in C programming language in a compact and modular design to facilitate its parallelization and optimization for modern HPC hardware.  The back-end kernel implements  stencil codes \cite{Datta2009Stencil} which are compiled and linked according to the hardware specified by the application programmer. Parameters provided by the front-end guide the generation of the back-end, which can implement either serial (baseline) or parallel code (in OpenMP or OpenACC), for 2D or 3D domains, to solve the acoustic wave with constant or variable density (Equations \ref{eq:acoustic_density} and \ref{eq:acoustic}, respectively), to execute on CPUs or GPUs. Once the back-end code is generated, it receives data structures initialized in the front-end and passed by parameters through Ctypes. The back-end executes the simulation and returns final results to the front-end.  

\indent 
The back-end supports all the concerns related to parallelism, performance, hardware specific optimizations and performance portability. Besides providing a reference  implementation which is numerically correct, the baseline (serial) code can also be used as an industry proxy of seismic applications for research in high performance computing (HPC) \cite{raut2020evaluating,zhou2021automated,michalowicz2021comparing,9651214,raut2021porting}. 
In its first release, {\tt simwave} implements three back-ends: sequential C (baseline), OpenMP, and OpenACC. The two later can generate code for CPUs of different architectures (e.g., x86, ARM, AMD, Power) and for GPUs.
In the future, novel back-ends may be developed  using technologies like OpenCL, DPC++, CUDA, and others. 

\section{Example of use}

\begin{figure}
    \centering
     \includegraphics[width=1.0\textwidth]{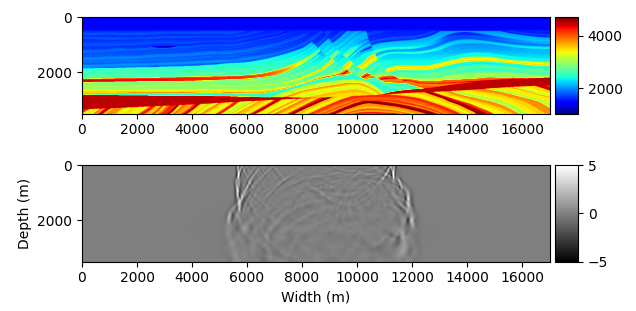}
    \caption{Marmousi2 velocity model (top) and final wavefield of the forward simulation (bottom).}
    \label{fig:marmousi_vel_model_u}
\end{figure}

This section illustrates the use of {\tt simwave} for the simulation of two examples. Listing~\ref{lst:marmousi_2d} shows the use of the {\tt simwave} to simulate acoustic waves propagation with the Marmousi2 P-wave velocity model (Figure~\ref{fig:marmousi_vel_model_u}) \cite{Gary2006Marmousi2} in a two dimensional domain which has 3.5 km depth by 17 km width. Other external packages (e.g. {\tt scipy}, {\tt matplotlib}, {\tt numpy}) can be used together for data visualization.

\begin{lstlisting}[language=Python, label={lst:marmousi_2d}, numbers=left, stepnumber=1, caption=Forward simulation in a two dimensional domain using Marmousi2 velocity model.]
from simwave import (
    SpaceModel, TimeModel, RickerWavelet, Solver, Compiler,
    Receiver, Source, read_2D_segy,
    plot_wavefield, plot_shotrecord, plot_velocity_model
)
import numpy as np

# Marmousi2 velocity model
marmousi_model = read_2D_segy('MODEL_P-WAVE_VELOCITY_1.25m.segy')

compiler = Compiler(
    cc='gcc',
    language='cpu_openmp',
    cflags='-O3 -fPIC -ffast-math -std=c99'
)

# create the space model
space_model = SpaceModel(
    bounding_box=(0, 3500, 0, 17000),
    grid_spacing=(10.0, 10.0),
    velocity_model=marmousi_model,
    space_order=4,
    dtype=np.float64
)

# config boundary conditions
space_model.config_boundary(
    damping_length=(0, 700, 700, 700),
    boundary_condition=(
        "null_neumann", "null_dirichlet",
        "null_dirichlet", "null_dirichlet"
    ),
    damping_polynomial_degree=3,
    damping_alpha=0.001
)

# create the time model
time_model = TimeModel(
    space_model=space_model,
    tf=2.0
)

# create the set of sources
source = Source(
    space_model,
    coordinates=[(20, 8500)],
    window_radius=1
)

# crete the set of receivers
receiver = Receiver(
    space_model=space_model,
    coordinates=[(20, i) for i in range(0, 17000, 10)],
    window_radius=1
)

# create a ricker wavelet with 10hz of peak frequency
ricker = RickerWavelet(10.0, time_model)

# create the solver
solver = Solver(
    space_model=space_model,
    time_model=time_model,
    sources=source,
    receivers=receiver,
    wavelet=ricker,
    saving_stride=0,
    compiler=compiler
)

# run the forward
u_full, recv = solver.forward()

# remove damping extension from u_full
u_full = space_model.remove_nbl(u_full)

extent = [0, 17000, 3500, 0]

# plot the velocity model
plot_velocity_model(space_model.velocity_model, extent=extent)

# plot the last wavefield
plot_wavefield(u_full[-1], extent=extent)

# plot the seismogram
plot_shotrecord(recv)
\end{lstlisting}

\indent After reading the velocity model (in line 9), we define the compiler options (lines 11-15) by instantiating an object {\tt Compiler}. This object defines a set of compiler choices and flags including the C compiler, the compilation flags, the {\tt language} which enables sequential or parallel implementation (in OpenMP or OpenACC), and the target architecture (i.e., CPU or GPU).
Optionally, it is possible to override the baseline code by pointing out to the path to a custom C implementation as kernel through the parameter {\tt cfile}. This can be useful to evaluate new strategies and HPC techniques. 

\indent Following this, we configure the spatial domain with the object {\tt SpaceModel} (line 18). The {\tt bounding\_box} attribute defines the domain boundaries (the begin and the end) in meters along the axis, respectively Z (depth) and X (width). 
The  {\tt grid\_spacing} defines the spacing (in meters) between grid points for each axis of the domain. The total grid size is calculated according to domain size ({\tt bounding\_box}) and the {\tt grid\_spacing}. The {\tt space\_order} defines the finite differences spatial order, which can be any even order ranging from 2 to 20. The {\tt dtype} sets the numeric precision, e.g.,  {\tt numpy.float32} for single-precision, and {\tt numpy.float64} for double-precision. The velocity model is represented as a numpy array in either two or three dimensions and expressed in meters per second by the {\tt velocity\_model} parameter. The optional attribute {\tt density\_model} specifies the density of materials in each grid point. When the density is provided, the acoustic equation with variable density (Equation \ref{eq:acoustic}) is auomtatically used for the simulation. In this case, the density model is also represented as a numpy array and carries the units of g/cm³. Both the velocity and density models are linearly interpolated to fit the domain extent.

\indent To enforce boundary conditions, line 27 invokes the method {\tt config\_boundary} of  {\tt SpaceModel}. The {\tt damping\_length} parameter defines the domain extension length (in meters) for the damping on each border of the domain, respectively Z (top and bottom) and X (left and right) in the 2D domain, and Y (front and back) in a 3D case. The parameter {\tt boundary\_condition} defines the boundary condition applied on each side of the domain. The options include {\tt null\_neumann}, {\tt null\_dirichlet}, and {\tt none}. The parameters {\tt damping\_polynomial\_degree} and {\tt damping\_alpha} are referred to as $p$ and $\alpha$ in Equation {\ref{eq:acoustic_damped}}.

\indent Next, (in line 38) we configure the time model by instantiating an object of the class  {\tt TimeModel}, providing the object {\tt space\_model} which contains spatial information (e.g. space order, domain dimension, maximum p-wave velocity) required to calculate the critical $\Delta t$. This object also encapsulates {\tt tf}, which defines total propagation time (in seconds) for the simulation, and {\tt saving\_stride} which sets the wave field saving configuration. The saving stride can be zero (only the snapshot in the last time step is returned), one (the snapshots of all time steps are returned) or any number $n$ ($1<n< \lceil \frac{tf}{\Delta t} \rceil$)  which determines saving every $n$ time steps (i.e., the stride). Optionally, the user  can define a custom $\Delta t$ through the optional parameter {\tt dt}, otherwise the default critical $\Delta t$ is applied.

\indent Next, we define the sources (in line 44) and receivers (line 51) by providing the grid (the {\tt space\_model} object), the  {\tt coordinates}, and {\tt window\_radius}. The parameter {\tt coordinates} is a list of tuples containing the coordinates of sources or receiver in meters in the domain. The {\tt window\_radius}  defines the radius (ranging from 1 to 10) of the Kaiser window applied in source/receiver interpolation. In the example, a Ricker wavelet is applied with a peak frequency of 10 Hertz. Notice that the wavelet requires the {\tt TimeModel}.

\begin{figure}[t]
    \centering
    \subfloat[Marmousi2.]{
        \includegraphics[scale=0.5]{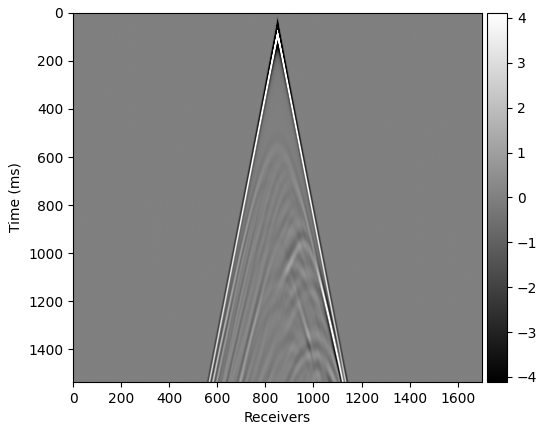}
        \label{fig:marmousi_shotrecord}
    }
    \subfloat[Overthrust.]{
        \includegraphics[scale=0.5]{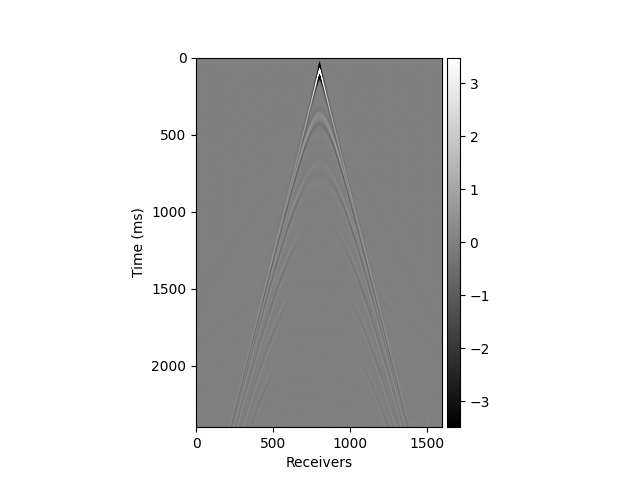}
        \label{fig:overthrust_shotrecord}
    }
    \caption{Seismogram from the the forward simulation.}
    \label{fig:seismogram}
\end{figure}

\indent An object  {\tt Solver} is instantiated by aggregating  all the previous objects that configure the simulation. The method  {\tt forward} executes the simulation, returning the full wave field and the seismogram after conclusion. The Figure~\ref{fig:marmousi_vel_model_u} shows the Marmousi2 velocity model (top) and the final wave field of the simulation (bottom), while Figure~\ref{fig:marmousi_shotrecord} depicts the corresponding seismogram.


\begin{figure}
    \centering
    \includegraphics[width=1.0\textwidth]{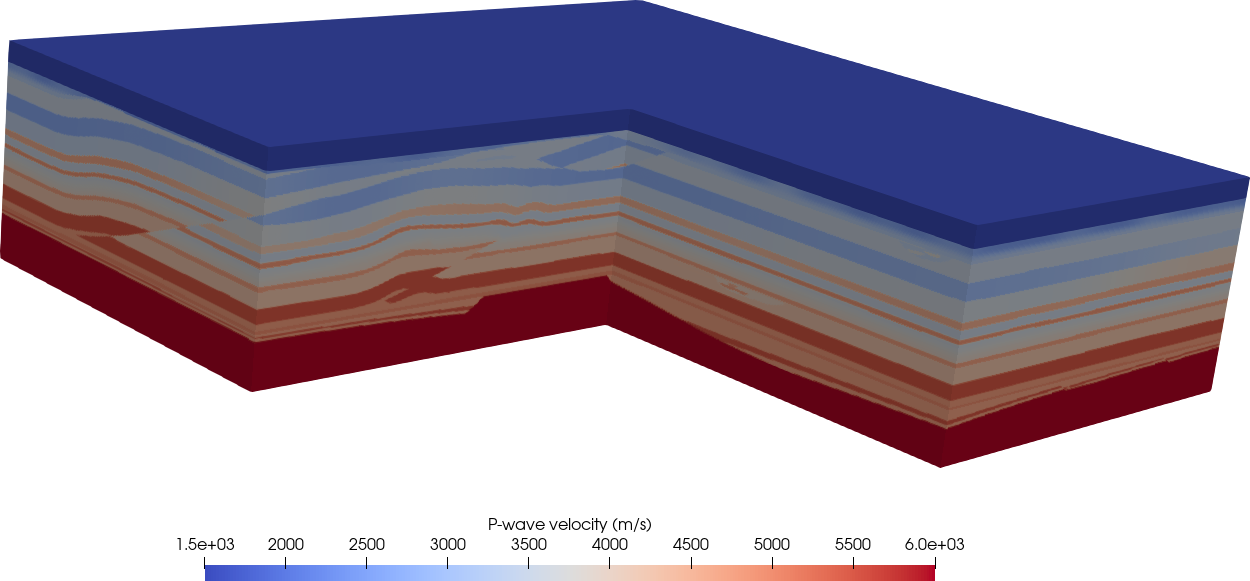}
    \caption{The Overthrust P-wave velocity model.}
    \label{fig:overthrust_vel_model_u}
\end{figure}

\begin{figure}
    \centering
    \includegraphics[width=1.0\textwidth]{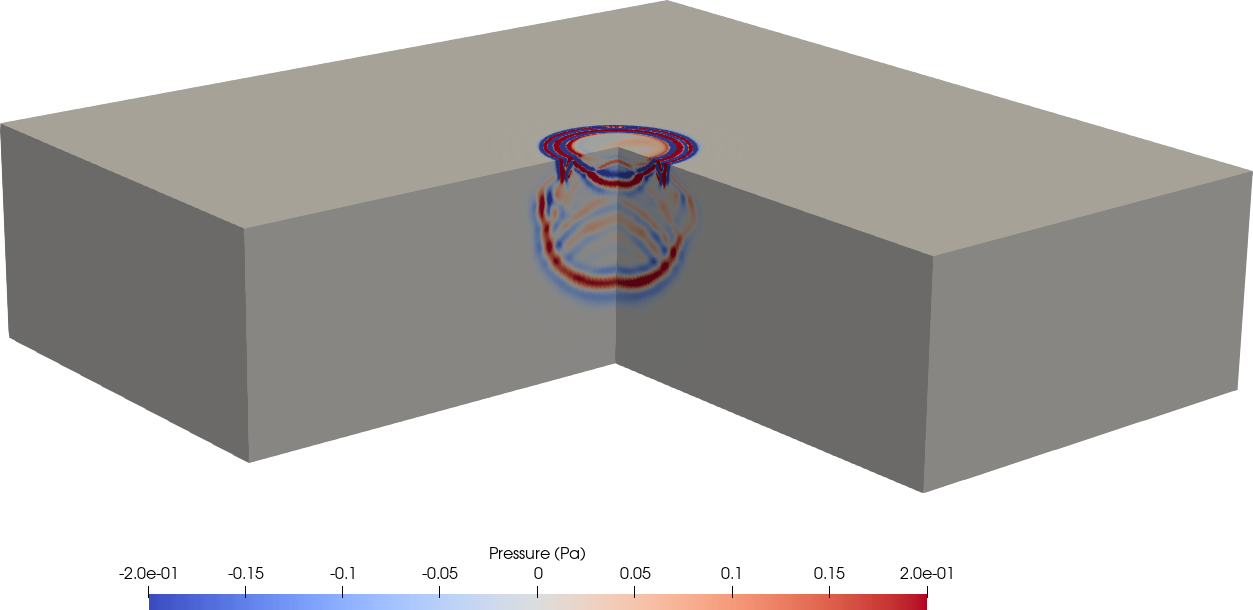}
    \caption{Wave field at $t=1.0$ s for the Overthrust benchmark.}
    \label{fig:overthrust_wave_field}
\end{figure}

\indent The next example (in Listing~\ref{lst:overthrust_3d}) shows the use of {\tt simwave} to simulate the propagation of an acoustic wave on the Overthrust velocity model  \cite{aminzadeh19973overthrust} in a three dimensional domain. The Overthrust model (depicted in Fig.~\ref{fig:overthrust_vel_model_u}) has 4.12 km in depth, 16 km in width and 16 km in length. The source code is very similar to the previous 2D example with the addition of one dimension. The final wave field produced by this simulation is shown in Fig. \ref{fig:overthrust_wave_field}, and the seismogram is shown in  Fig.~\ref{fig:overthrust_shotrecord}. 

\begin{lstlisting}[language=Python, label={lst:overthrust_3d}, numbers=left, stepnumber=1,  caption=Forward simulation in a three dimensional domain using the Overthrust velocity model.]
from simwave import (
    SpaceModel, TimeModel, RickerWavelet, Solver,
    Compiler, Receiver, Source
)
import numpy as np
import h5py

def read_model(filename):
    with h5py.File(filename, "r") as f:

        # Get the data
        data = list(f['m'])

        data = np.array(data)

        # convert to m/s
        data = (1 / (data ** (1 / 2))) * 1000.0

        return data

if __name__ == '__main__':

    data = read_model('overthrust_3D_true_model.h5')

    compiler = Compiler(
        cc='clang',
        language='gpu_openmp',
        cflags='-O3 -fPIC -ffast-math -fopenmp \
               -fopenmp-targets=nvptx64 -Xopenmp-target'
    )

    space_model = SpaceModel(
        bounding_box=(0, 4120, 0, 16000, 0, 16000),
        grid_spacing=(20., 20., 20.),
        velocity_model=data,
        space_order=4,
        dtype=np.float64
    )

    space_model.config_boundary(
        damping_length=0,
        boundary_condition=(
            "null_neumann", "null_dirichlet",
            "null_dirichlet", "null_dirichlet",
            "null_dirichlet", "null_dirichlet"
        ),
        damping_polynomial_degree=3,
        damping_alpha=0.001
    )

    time_model = TimeModel(
        space_model=space_model,
        tf=4,
        saving_stride=0
    )

    source = Source(
        space_model,
        coordinates=[(20, 8000, 8000)],
        window_radius=1
    )

    receiver = Receiver(
        space_model=space_model,
        coordinates=[(40, 8000, i) for i in range(0, 16000, 10)],
        window_radius=1
    )

    ricker = RickerWavelet(8.0, time_model)

    solver = Solver(
        space_model=space_model,
        time_model=time_model,
        sources=source,
        receivers=receiver,
        wavelet=ricker,
        compiler=compiler
    )

    # run the forward
    u_full, recv = solver.forward()
\end{lstlisting}

\section{Performance evaluation}

As wave simulation is the kernel of many large inversion problems, optimizing its performance for efficient execution on several HPC systems is mandatory. The two previous examples are used to assess performance on CPU and GPU systems. For the 2D performance test, a 2 second acoustic wave propagation using the Marmousi2 (Listing~\ref{lst:marmousi_2d}) benchmark in which a 3.5 km deep per 17 km wide domain is discretized with a 2D grid with 351 x 1701 points. A damping length of 700 m is added on each side except along the top boundary and results in a 421 x 1841 grid (775,061 grid points). The number of timesteps varies according to the spatial order, being 1331 for $2^{nd}$,  1537 for $4^{th}$, and 1696 for $8^{th}$. For the 3D performance experiment, we simulated 4 seconds of wave propagation using the Overthrust 3D velocity model (Listing~\ref{lst:overthrust_3d})  with 4.12 km depth x 16 km width x 16 km length, discretized in a 3D grid with 207 x 801 x 801 points (132,811,407 grid points). Likewise, the number of timesteps varies in 2080 for $2^{nd}$,  2401 for $4^{th}$ and 2651 for $8^{th}$ spatial order.

\indent Benchmarks were executed in both CPU and GPU environments. The CPU execution was carried out in a cluster node with two Intel Xeon Gold 6148 processors (Skylake) with 20 cores each and 192GB of memory. The GPU executions were performed in the GeForce RTX 2080 Super (Turing architecture) and Tesla V100 (Volta architecture). Each benchmark was compiled with GCC 8.3 (GNU Compiler) in the CPU environment. For the GPU execution we used the PGCC 21.11 (PGI compiler) for offloading using OpenACC and CLANG 13.0 (LLVM project)
for OpenMP. The flags applied in each compiler are listed in Table \ref{tab:flags}. 

\begin{table}[ht]
    \centering
    \caption{Compiler flags used in the executions.} 
    \label{tab:flags}
    \begin{tabularx}{\textwidth}{lX}
    \toprule
    Compiler & Flags \\
    \hline
    GCC & {\tt -O3 -fPIC -ffast-math -std=c99} \\
    CLANG & {\tt -O3 -fPIC -ffast-math -fopenmp -fopenmp-targets=nvptx64 -Xopenmp-target} \\
    PGCC & {\tt -O3 -fPIC -acc:gpu -gpu=pinned} \\
    \bottomrule
\end{tabularx}
\end{table}

\indent The experiment measured the execution time for both 2D and 3D acoustic wave propagator with constant density, discretized with $2^{nd}$, $4^{th}$, and $8^{th}$ spatial orders. For the CPU experiments we increased the number of cores from 1 core in the sequential version up to 40 cores available in the compute node. The execution on GPUs used all the available cores. 

\begin{table}[ht]
   \centering
   \caption{Execution times in seconds and speedup (parallel time / serial time) for the simulation of 2D Marmousi forward propagation. The best results are highlighted in bold.} \label{tab:performance2D}
   \resizebox{\textwidth}{!}{%
   \begin{tabular}{ccccc cc cc}
   \toprule
   \multirow{2}{*}{Hardware} & \multirow{2}{*}{Back-end} &
   \multirow{2}{*}{Compiler} &
   \multicolumn{2}{c}{SO=2} & \multicolumn{2}{c}{SO=4} &  \multicolumn{2}{c}{SO=8} \\
   \cmidrule(l){4-9}
   & & & Time & S & Time & S & Time & S \\
\hline
6148 - 1 core & C & gcc &4.18 & 1.0 & 6.27 & 1.0 & 10.03 & 1.0 \\
6148 - 2 cores & OpenMP & gcc& 2.31 & 1.8 & 3.40 & 1.8 & 5.38 & 1.9 \\
6148 - 4 cores & OpenMP & gcc& 1.27 & 3.3 & 1.87 & 3.4 & 2.90 & 3.5 \\
6148 - 8 cores & OpenMP & gcc& 0.71 & 5.9 & 1.04 & 6.0 & 1.57 & 6.4 \\
6148 - 20 cores & OpenMP & gcc& 0.39 & 10.7 & 0.57 & 11.1 & 0.83 & 12.0 \\
6148 - 40 cores & OpenMP & gcc& {\bf 0.31} & {\bf 13.3} & 0.48 & 13.1 & 0.69 & 14.5 \\
RTX 2080 Super & OpenMP & clang& 0.76 & 5.5 & 0.86 & 7.3 & 0.98 & 10.3 \\
RTX 2080 Super & OpenACC & pgcc& 0.61 & 6.9 & 0.66 & 9.5 & 0.77 & 13.0 \\
V100 & OpenMP & clang & 0.36 & 11.5 & {\bf 0.43} & {\bf 14.6} & {\bf 0.47} & {\bf 21.3} \\
V100 & OpenACC &pgcc& 0.41 & 10.1 & 0.45 & 14.1 & 0.49 & 20.5 \\
\bottomrule
\end{tabular}
}
\end{table}

\indent The simulations were repeated 10 times and the average execution times and speedups are presented in the Tables \ref{tab:performance2D} (Marmousi 2D) and \ref{tab:performance3D} (Overthrust 3D). Note that the speedup is calculated as the ratio of the parallel execution time to the serial execution time. 
Because such finite difference stencils are intrinsically memory-bound codes, the scalability in CPU is hindered when the number of cores is increased above the number of memory channels available (6 channels for this CPU). This result is consistent with other studies in literature (e.g., in \cite{Micikevicius2009,Pershin2019}.

\indent For the 2D benchmark with spatial order 2, the CPU with 40 cores produces the best performance. However, for higher spatial orders, the GPU performs better. In the case of the GPU, there is a data transfer cost (CPU memory to GPU memory), but because the GPU has far higher throughput in terms of processing than the CPU, our results suggest larger workloads can be processed on the GPU more quickly. Further, the numerical solution of the wave equation implements stencil patterns which is memory-intensive, a scalability limiting factor \cite{SCHAFER2011Stencil,Micikevicius2009}. Thus, the memory bandwidth represents a bottleneck for performance and scalability. 

\indent Notice that 3D benchmark showed speedups significantly higher than the 2D because the 3D produces a significantly larger amount of work to execute. By calculation 132,811,407 grid points per time step, the 3D launches a massive number of work units (i.e., thread blocks) which can be executed in parallel as soon as their data arrive from the memory. This allows better hiding the memory latency of the GPU than the 2D benchmark which computes far less (775,061) grid points per time step.

\indent Currently, {\tt simwave} applies straightforward loop parallelism strategies supported by thread-based OpenMP or OpenACC compilers, and compiler-specific automatic optimizations (i.e. -O3). The investigation on more advanced loop optimization strategies is beyond of this work's scope and will be addressed in future work.

\begin{table}[ht]
   \centering
   \caption{Execution times (in seconds) and speedup for the simulation of 3D Overthrust forward propagation. The best results are highlighted in bold.} 
   \label{tab:performance3D}
   \resizebox{\textwidth}{!}{%
   \begin{tabular}{ccccc cc cc}
   \toprule
   \multirow{2}{*}{Hardware} & \multirow{2}{*}{Back-end} &
   \multirow{2}{*}{Compiler} &
   \multicolumn{2}{c}{SO=2} & \multicolumn{2}{c}{SO=4} &  \multicolumn{2}{c}{SO=8} \\
   \cmidrule(l){4-9}
   & & & Time & S & Time & S & Time & S \\
\hline
6148 - 1 core & C & gcc & 1642.41 & 1.0 & 2565.88 & 1.0 & 3909.55 & 1.0 \\ 
6148 - 2 cores & OpenMP & gcc & 901.51 & 1.8 & 1360.49 & 1.9 & 2048.54 & 1.9 \\ 
6148 - 4 cores & OpenMP & gcc & 475.52 & 3.5 & 716.31 & 3.6 & 1081.00 & 3.6 \\
6148 - 8 cores & OpenMP & gcc & 248.84 & 6.6 & 374.98 & 6.8 & 569.71 & 6.9 \\ 
6148 - 20 cores & OpenMP & gcc & 186.98 & 8.8 & 272.56 & 9.4 & 429.54 & 9.1 \\ 
6148 - 40 cores & OpenMP & gcc & 110.72 & 14.8 & 171.11 & 15.0 & 347.09 & 11.3 \\ 
RTX 2080 Super & OpenMP & clang & 72.46 & 22.7 & 93.95 & 27.3 & 130.23 & 30.0 \\
RTX 2080 Super & OpenACC & pgcc & 48.02 & 34.2 & 67.68 & 37.9 & 103.95 & 37.6 \\
V100 & OpenMP & clang & {\bf 28.30}&{\bf 58.0}&{\bf40.36}&{\bf 63.6}&{\bf 63.12} &{\bf 61.9} \\
V100 & OpenACC & pgcc & 37.13 & 44.2 & 50.10 & 51.2 & 68.13 & 57.4 \\
\bottomrule
\end{tabular}
}
\end{table}

\section{Comparison with other simulation packages}




{\tt simwave} implements an explicit solver to simulate the propagation of acoustic waves with constant or variable density, based on the finite-difference method. 
There are plenty of software technologies used for Geophysics research, including those developed by communities \cite{CIGmisc, VerceProject}, companies, or individuals \cite{wiki:ComparisonFreeGeophysicsSoftware}.
However, most software packages maintained by the CIG project \cite{CIGmisc} are not directly comparable to {\tt simwave} because they were designed with focus on specific aspects of earthquakes.

\indent A comprehensive list in \cite{wiki:ComparisonFreeGeophysicsSoftware} compares dozens of software packages which focus on exploration geophysics.
Likely the most widely known software for geophysics research,  
Madagascar \cite{home:madagascar}  is designed for multidimensional data analysis and reproducible computational experiments which is distributed as an open-source package \cite{madagascar_github}. 
The objective is to provide an environment for researchers working with digital image and data processing in geophysics and related fields. The package consists of two levels: low-level main programs (typically developed in the C programming language and working as data filters) and high-level processing flows (described with the help of the Python programming language) that combine main programs and completely document data processing histories for testing and reproducibility. The package is composed of more than 1,000 programs that support a significantly broader range of functionalities if compared to {\tt simwave}. Furthermore, Madagascar's focus is to serve as a tool for reproducible research in several areas of geophysics, while {\tt simwave} focus on simulate the propagation of acoustic waves.  

\indent A more closely related project is Minimod \cite{meng2020minimod}, which implements several solvers for the acoustic wave with constant density, acoustic wave with variable density, acoustic transversely isotropic, and the elastic equation. Minimod can serve both for geophysics research, and for HPC research as in \cite{raut2020evaluating,zhou2021automated,michalowicz2021comparing,9651214,raut2021porting}. However, Minimod is currently not publicly available by the time of this writing. 

\section{Quality control}


Quality control is enforced with the support of {\tt pytest} (\url{https://docs.pytest.org/}), and continuous integration and continuous delivery (CI/CD) mechanisms supported by GitHub. 
This enables automating the tests and running the software development workflows directly in the repository using the GitHub's servers. Tests are executed on every push or pull requests to the master branch of the repository.
Similarly, one workflow uploads and updates the simwave's package version in the PyPI every time a release is created. 
The test suite consists of functional and unit tests. 
The unit tests are important to check isolated pieces of the code, ensuring the expected outputs according to the inputs. And the functional tests are applied to verify slices of the application as well as the entire program. Part of the tests are black box, comparing the simulation output to known reference values. These values are obtained from problems that have analytical solutions (described in the validation section) and also from the output of earlier versions of simwave, which is a form of regression testing. Installation and testing instructions can be found in the {\tt simwave}'s repository on GitHub, along with use case examples.

\section{Availability}
\vspace{0.5cm}
\section*{Operating system}
{\tt simwave} can be installed via {\tt pip} package manager either from the source or from the Python Package Index (PyPI) on GNU/Linux, Mac OS X and on any platform supported by Docker, like Azure and AWS. 

\section*{Programming language}
Python 3.6 or newer and C.

\section*{Additional system requirements}
Memory depending on domain size and use case.

\section*{Dependencies}
The required {\tt simwave} dependencies are listed below.

\begin{enumerate}
    \item numpy$>=$1.18.1  
    \item matplotlib$>=$3.2.1
    \item segyio$>=$1.9.1
    \item scipy$>=$1.4.1
    \item pytest$>=$6.2.2
    \item pytest-codeblocks$>=$0.10.4
    \item findiff$>=$0.8.9   
 \end{enumerate}

\section*{List of contributors}
The development of {\tt simwave} has the following contributors.

\begin{enumerate}
    \item Jaime Freire de Souza (Federal University of Sao Carlos)
    \item Keith Jared Roberts (University of Sao Paulo)
    \item João Baptista Dias Moreira (University of Sao Paulo)
    \item Roussian di Ramos Alves Gaioso (Federal University of Sao Carlos)
    \item Hermes Senger (Federal University of Sao Carlos)
 \end{enumerate}

\section*{Software location:}

{\bf Archive} 

\begin{description}[noitemsep,topsep=0pt]
	\item[Name:] Zenodo 
	\item[Persistent identifier:] \url{https://doi.org/10.5281/zenodo.5847017}
	\item[Licence:] GNU General Public License v3.0 
	\item[Publisher:] Hermes Senger 
	\item[Version published:] v1.0 
	\item[Date published:] {13/01/2022}
\end{description}

{\bf Code repository} 

\begin{description}[noitemsep,topsep=0pt]
	\item[Name:] GitHub
	\item[Persistent identifier:] \url{https://github.com/HPCSys-Lab/simwave}
	\item[Licence:] GNU General Public License v3.0
	\item[Date published:] 13/01/2022
\end{description}



\section*{Language}
English.

\section*{Reuse potential}


{\tt simwave} can be used to simulate the propagation of acoustic waves in single- or multi-material domains with constant and variable density, such as in full-waveform inversion (FWI) \cite{virieux2009overview} or reverse-time migration (RTM) \cite{baysal1983reverse,tarantola1984inversion} problems.
The simulations are written in Python and use {\tt simwave} as a library to be imported and used either alone, or in combination with scientific libraries such as SciPy and others.
The {\tt simwave}'s code is provided in two forms, a sequential (baseline) and an accelerated implementation for users who need to cope with large problems. The code may also be used as a representative of relevant industrial codes which can serve as benchmark for research on high-performance computing methods, such as in \cite{raut2020evaluating,zhou2021automated,michalowicz2021comparing,9651214,raut2021porting}. 
Finally, users and researchers can get in touch with the development team through the {\tt simwave}’s issue page on GitHub (\url{https://github.com/HPCSys-Lab/simwave/issues}).

\section*{Acknowledgements}


The authors thank the support  from Shell Brasil and ANP. The sixth author thanks the financial support of CNPq, Brazil under grant 302658/2018-1. The seventh author thanks the support of São Paulo Research Foundation (FAPESP), under grant 2019/26702-8. 

\section*{Funding statement}

The authors gratefully acknowledge sponsorship from Shell Brasil through the {\it ANP 20714-2 - Desenvolvimento de técnicas numéricas e software para problemas de inversão com aplicações em processamento sísmico} project at Universidade de São Paulo and the strategic importance of the support given by ANP through the R\&D levy regulation.

\section*{Competing interests}

The authors have no competing interests to declare.




\bibliographystyle{agsm}

\bibliography{99.sample,99.related-work}

\vspace{2cm}

\rule{\textwidth}{1pt}





\end{document}